\documentclass[twocolumn,aps]{revtex4}

\usepackage[english]{babel}
\usepackage[makeroom]{cancel}
\usepackage[utf8]{inputenc}
\usepackage{amsmath}
\usepackage{amssymb}
\usepackage{amstext}
\usepackage{braket}
\usepackage{graphicx}
\usepackage{hyperref}
\usepackage{bbm}

\usepackage{qcircuit}
\usepackage{tikz}
\usepackage{tikz-cd}
\usetikzlibrary{decorations.markings,patterns,shapes.geometric}
\usepackage{ifthen}

\graphicspath{{.}{abrak/Bloch/}{abrak/p_i_filled/}}
\DeclareGraphicsExtensions{.png,.jpg}

\usepackage{epsfig}
\usepackage[labelformat=simple]{subcaption}
\usepackage{floatrow}

\usepackage{ragged2e}

\usepackage{comment}

\newif\ifcount
\countfalse

\newcommand\mywordcount{
\ifcount
	$ \phantom{\sum}$ \\ \noindent\textbf{\large Wordcount} \\ $\phantom{\sum}$ \\
	\noindent\input|"texcount -sub=section \jobname.tex | grep -e Words -e Number -e Section -e top -e Part | awk 1 ORS='\string\\\string\\' | sed -e 's/\string\_/ /g'"
	text+headers+captions (\#headers/\#floats/\#inlines/\#displayed)\\
\else
\fi
}

\newcommand{\ourappendix}{Appendix}

\newcommand{\mqquad}{\!\!\!\!\!\!\!\!}
\newcommand{\mqqquad}{\mqquad\mqquad}
\newcommand{\torus}{
\scalebox{1}{
\begin{tikzpicture}[scale=0.1]
\draw[cm={1,0,0,1,(0cm,-0cm)}] (0,0) ellipse (3.6cm and 1.8cm);

\draw[cm={1,0,0,1,(0cm,-0cm)}] (-2,.3) .. controls (-1.5,-0.2) and (-1,-0.4) .. (0,-.4) .. controls (1,-0.4) and (1.5,-0.2) .. (2,0.3);

\draw[cm={1,0,0,1,(0cm,-0cm)}] (-1.65,0) .. controls (-1.5,0.3) and (-1,0.5) .. (0,.5) .. controls (1,0.5) and (1.5,0.3) .. (1.65,0);
\end{tikzpicture}
}
}
\newcommand{\boxedtorus}{\mbox
  {\vbox to .25cm {\vfil \vskip -0.050cm
  \hbox to 0.76cm
  {\hskip -0.13cm $\torus$}
  \vfil
}}
}

\newcommand{\sphere}{
\bigcirc
}

\begin{document}
\title{Exponential Sensitivity and its Cost in Quantum Physics}

\author{András Gilyén$^{1,3}$}
\email{gilyen.andras@wigner.mta.hu}

\author{Tamás Kiss$^1$}
\email{kiss.tamas@wigner.mta.hu}

\author{Igor Jex$^2$}

\affiliation{$^1$ Institute for Solid State Physics and Optics, Wigner Research Centre for Physics, Hungarian Academy of Sciences, Konkoly-Thege M. u. 29-33, H-1121 Budapest, Hungary}
\affiliation{$^2$ Faculty of Nuclear Sciences and Physical Engineering, Czech Technical University in Prague,
B\v{r}ehová 7, 115 19 Praha 1-Staré M\v{e}sto, Czech Republic}
\affiliation{$^3$ Department of Theoretical Physics, Budapest University of Technology and Economics, Budafoki út 8, H-1111 Budapest, Hungary}


\begin{abstract}
State selective protocols, like entanglement purification, lead to an essentially non-linear quantum evolution, unusual in naturally occurring quantum processes. Sensitivity to initial states in quantum systems, stemming from such non-linear dynamics, is a promising perspective for applications. Here we demonstrate that chaotic behaviour is a rather generic feature in state selective protocols: exponential sensitivity can exist for all initial states in an experimentally realisable optical scheme. Moreover, any complex rational polynomial map, including the example of the Mandelbrot set, can be directly realised. In state selective protocols, one needs an ensemble of initial states, the size of which decreases with each iteration. We prove that exponential sensitivity to initial states in any quantum system have to be related to downsizing the initial ensemble also exponentially. Our results show that magnifying initial differences of quantum states (a Schrödinger microscope) is possible, however, there is a strict bound on the number of copies needed.
\end{abstract}

\maketitle

\vskip 4 mm \noindent \textbf{\large Introduction} \\* \noindent
Quantum technology progresses at a fast pace. Preparation, control and measurement of coherent quantum systems \cite{QControl} became possible on an unprecedented level leading to a wealth of proposals of applications ranging from quantum information processing to high precision measurements and sensors. In these protocols, increasingly sophisticated sequences of coherent evolution, measurement and post-selection are applied in order to control the state of quantum systems. Dynamics achieved by state selective protocols was proven essential for a large number of quantum information protocols \cite{KLM,Calsamiglia} and quantum communication \cite{QRepeater}. Prominent examples of probabilistic protocols are the KLM scheme \cite{KLM} for linear optical quantum gates or the entanglement purification protocols \cite{purification1,purification2,purification3} employing measurement and selection in order to increase the entanglement between subsystems.

Manipulation by measurement and selection breaks the linearity of quantum mechanics, thereby broadens the possibilities for quantum evolution \cite{Gisin,Scott,Habib,Everitt}. 
In contrast in the well established field of quantum chaos \cite{ChaosBook} one studies the signatures of chaos in closed quantum systems with linear evolution.
However, the essential non-linearity of an iterated, state selective protocol can result in truly chaotic behaviour, showing exponential sensitivity to initial conditions \cite{OurPRA2006,2011,Guan}. So far sensitivity has been proved only for a tiny fractal subset of initial states with zero measure. In this article we demonstrate that exponential sensitivity can exist for all initial states in an experimentally realisable optical scheme. Moreover, we show that any complex rational polynomial map, including the example of the Mandelbrot maps \cite{DevaneyBook}, can be directly realised using state selective protocols bringing a whole new class of quantum protocols to life.

From a fundamental point of view, one can search for the most general evolution for a quantum system. A very general dynamics is sometimes imagined as a system together with one or more ancillas and allowing for both unitary evolution and non-selective measurements on the complete arrangement. The evolution reduced for the system only is called a quantum channel. When talking about quantum states in practice, it is unavoidable to be able to repeat experiments on an ensemble of identically prepared initial states, in order to uncover the underlying probabilistic laws. This ensemble view of quantum states allows for the following trick when designing the most general dynamics for a given initial state. Let us, for example, consider systems from the ensemble pairwise and let them interact with each other. After the interaction one can perform a measurement on one of the pairs and then discard the measured member of the pair. In case of selective measurement, one may also discard the unmeasured member of the pair, depending on the measurement result.  The resulting ensemble will be reduced in size, but some of its properties may be changed in a beneficial way, e.g. entanglement between subsystems. The above procedure goes beyond the usual notion of quantum channels, in the following sense. The initial step of the procedure, namely taking the systems pairwise, can be viewed as splitting the original ensemble into two parts, and employing one part as an ancilla. In other words, the state of the ancilla will be dependent on the state of the system. The state dependent ancilla lies at the heart of the non-linearity of the process. 

$\displaystyle{\phantom{\sum}}$

$\displaystyle{\phantom{\sum}}$

\vskip 4 mm \noindent \textbf{\large Results} \\* 
\noindent \textbf{A linear optical experimantal scheme implementing a family of non-linear maps.}
We propose here a simple experimental setup which implements a non-linear process exhibiting exponential sensitivity to the initial state.
Our scheme is inspired by an experimentally tractable entanglement purification protocol  \nolinebreak\cite{PBS,PBSExperiment} and uses only linear optical elements.  
During iterations we form pairs of photonic qubits from an ensemble of identically prepared photons and apply a post-selective transformation on the pairs by measuring the polarization of one photon and keeping or throwing away the other photon depending on the measurement result. The post selection induces a non-linear, deterministic transformation on the remaining photons, therefore the kept photons remain identically prepared.

\colorlet{maroon}{red!80!black}
\newcommand{\PBS}[8]{
	\ifthenelse{#2=1}
	{
		\draw[thick,maroon, postaction=decorate, decoration={markings,
			mark= at position 0.29 with {\arrow[rotate=0]{stealth};},
			mark= at position 0.79 with {\arrow[rotate=0]{stealth};}
		}] (-1,1) -- (0,0) -- (1,-1);
		\node[label=above:{#3}] (L1) at (-0.5,  0.5) {} ;
		\node[label=right:{#4}] (L2) at ( 0.5, -0.5) {} ;		
	}
	{
		\draw[thick,maroon, postaction=decorate, decoration={markings,
			mark= at position 0.29 with {\arrow[rotate=0]{stealth};},
			mark= at position 0.79 with {\arrow[rotate=0]{stealth};}
		}] (-1,1.08) -- (0,0.08) -- (1,1.08);
		\node[label=above:{#3}] (L1) at (-0.5,  0.5) {} ;
		\node[label=above:{#4}] (L2) at ( 0.5,  0.5) {} ;	
	}
	\ifthenelse{#5=1}
	{
		\draw[thick,maroon, postaction=decorate, decoration={markings,
			mark= at position 0.29 with {\arrow[rotate=0]{stealth};},
			mark= at position 0.79 with {\arrow[rotate=0]{stealth};}
		}] (-1,-1) -- (0,0) -- (1,1);
		\node[label=below:{#6}] (L1) at (-0.5, -0.5) {} ;
		\node[label=right:{#7}] (L2) at ( 0.5,  0.5) {} ;		
	}
	{
		\draw[thick,maroon, postaction=decorate, decoration={markings,
			mark= at position 0.29 with {\arrow[rotate=0]{stealth};},
			mark= at position 0.79 with {\arrow[rotate=0]{stealth};}
		}] (-1,-1.08) -- (0,-0.08) -- (1,-1.08);
		\node[label=below:{#6}] (L1) at (-0.5, -0.5) {} ;
		\node[label=below:{#7}] (L2) at ( 0.5, -0.5) {} ;	
	}	
	\draw node[style={rectangle,draw,fill=blue,minimum size=10mm,thin},fill opacity=0.25,rotate=-45] (N) {};
	\draw[thick,rotate around={30:(-1,0.5)}] (N.south west)--(N.north east);
	\node[label=left:{#1}] (PBS) at (N.south west) {};
	
	\ifthenelse{#8=0}{}
	{
		\draw node[style={rectangle,draw,fill=yellow,minimum size=6mm,thin},fill opacity=0.25,rotate=-45] (U) at (1,#8) {};
		\node at (U) {U};	
	
		\ifthenelse{#8=1}	
		{\draw node[style={rectangle,draw,minimum size=6mm,thin},rotate=-45, pattern= vertical lines 
		, pattern color=black] (M) at (1,0-#8) {};}
		{\draw node[style={rectangle,draw,minimum size=6mm,thin},rotate=-45, pattern= vertical lines 
		, pattern color=black] (M) at (1,0-#8) {};}
		\node (O) at (0,0){};
		\node (R) at (2,0-#8-#8){};
		\draw[thin,maroon,] (barycentric cs:O=0.5,R=0.5) -- (barycentric cs:O=0.3,R=0.7);
		\node[shape=semicircle,rotate=-90-\the\numexpr#8*45,fill=gray,fill opacity=0.75,inner sep=3pt, anchor=south, outer sep=0mm] (D) at (barycentric cs:O=0.325,R=0.675) {}; 
	}	
}

\newcommand{\scalePBS}{0.7}
\begin{figure}[ht!]
	\phantomsection
	\centering
	\begin{subfigure}{.2\textwidth}
	  \centering
	  \scalebox{\scalePBS}{
	  \begin{tikzpicture}
		  \PBS{}{1}{1}{4}{1}{2}{3}{0}
	  \end{tikzpicture}
	  }
	  \caption{PBS}
	  \label{fig:1234}
	\end{subfigure}%
	\begin{subfigure}{.2\textwidth}
	  \centering
  	  \scalebox{\scalePBS}{
	  \begin{tikzpicture}
	  	\PBS{}{1}{H}{H}{1}{H}{H}{0}
	  \end{tikzpicture}
	  }
	  \caption{$\!\!\ket{H}_{\!1}\!\ket{H}_{\!2}\!\!$}
	  \label{fig:HH}
	\end{subfigure}%
	\begin{subfigure}{.2\textwidth}
	  \centering
	  \scalebox{\scalePBS}{
	  \begin{tikzpicture}
	  	\PBS{}{-1}{V}{V}{-1}{V}{V}{0}
	  \end{tikzpicture}
	  }
	  \caption{$\!\!\ket{V}_{\!1}\!\ket{V}_{\!2}\!\!$}
	  \label{fig:VV}
	\end{subfigure}%
	\begin{subfigure}{.2\textwidth}
	  \centering
	  \scalebox{\scalePBS}{
	  \begin{tikzpicture}
	  	\PBS{}{1}{H}{H}{-1}{V}{V}{0}
	  \end{tikzpicture}
	  }
	  \caption{$\!\!\ket{H}_{\!1}\!\ket{V}_{\!2}\!\!$}
	  \label{fig:HV}
	\end{subfigure}%
	\begin{subfigure}{.2\textwidth}
	  \centering
	  \scalebox{\scalePBS}{
	  \begin{tikzpicture}
	  	\PBS{}{-1}{V}{V}{1}{H}{H}{0}
	  \end{tikzpicture}
	  }
	  \caption{$\!\!\ket{V}_{\!1}\!\ket{H}_{\!2}\!\!$}
	  \label{fig:VH}
	\end{subfigure}%
	\\
	\vskip 0.3 cm
	\ffigbox[]{ 
		\TopFloatBoxes
		\begin{subfloatrow}
			\vtop to72mm
			{
				\floatsetup[figure]{floatrowsep=none} 
				\killfloatstyle
				\par
				\ffigbox[.8\hsize]
				{\caption{Full scheme}\label{fig:LinFull}}
				{
					\scalebox{0.8}{
					\begin{tikzpicture}
						\begin{scope}[yshift=0cm]
							\PBS{}{1}{}{}{1}{}{}{0}
							
							\draw (0.95,0.63) -- (0.95,-0.63);
							\draw (1.05,0.63) -- (1.05,-0.63);
						
							\draw[thick,maroon, postaction=decorate, decoration={markings,
								mark= at position 0.6 with {\arrow[rotate=0]{stealth};}
							}] (1,1) -- (2.15,2.15);		
								
							\draw node[style={rectangle,draw,fill=gray,minimum size=6mm,thin},fill opacity=0.25,rotate=-45] (M) at (1,-1) {};
							\draw[thick,rotate around={30:(-1,0.5)}] (M.south east)--(M.north west);
							\node at (M) {$+ \, -$};
							\node[shape=semicircle,rotate=-135,fill=gray,fill opacity=0.75,inner sep=3pt, outer sep=0mm,minimum size=3.0mm,] (DPM) at (1.3,-1.3) {}; 
							
							\draw node[style={rectangle,draw,fill=yellow,minimum size=6mm,thin},fill opacity=0,rotate=-45] (Z) at (1,1) {};
							\node at (Z) {Z};		
						\end{scope}
					\end{tikzpicture}
					}
				}
				
				\vss
				\ffigbox[.8\hsize]
				{\caption{Simplified scheme}\label{fig:LinSimple}}
				{
					\scalebox{0.8}{
					\begin{tikzpicture}
						\begin{scope}[yshift=0.0cm]
							\PBS{}{1}{}{}{1}{}{}{0}
							\draw node[style={rectangle,draw,minimum size=6mm,thin},rotate=-45, pattern= vertical lines, pattern color=black] (M) at (1,-1) {};
							\node (O) at (0,0){};
							\node (R) at (2,-2){};
							\draw[thin,maroon,] (barycentric cs:O=0.5,R=0.5) -- (barycentric cs:O=0.3,R=0.7);
							\node[shape=semicircle,rotate=-135,fill=gray,fill opacity=0.75,inner sep=3pt, anchor=south, outer sep=0mm] (D) at (barycentric cs:O=0.325,R=0.675) {}; 
						\end{scope}	
					\end{tikzpicture}
					}
				}
							
				\vskip0cm
			}
			
			\ffigbox[\Xhsize][72mm]
			{\caption{Two iterations}\label{fig:LinTwoScheme}}
			{
				\scalebox{0.8}{
				\begin{tikzpicture}
					\begin{scope}[xshift=2cm,yshift=-2cm]
						\draw[thick,maroon,] (1,1) -- (1.5,1.5);
						
						\draw node[style={rectangle,draw,fill=gray,minimum size=6mm,thin},fill opacity=0.25,rotate=-45] (FHV) at (1.5,1.5) {};
						\draw[thick,rotate around={30:(-1,0.5)}] (FHV.south west)--(FHV.north east);
						\node at (FHV) {$\!\!\!\!\!\!\phantom{\sum}^{\text{H}}_{\text{V}}$};
						\node[shape=semicircle,rotate=-45,fill=gray,fill opacity=0.75,inner sep=3pt, outer sep=0mm,minimum size=3.0mm,] (DPM) at (1.8,1.8) {}; 	
						\PBS{}{1}{}{}{1}{}{}{1}
					\end{scope}		
					\begin{scope}
						\PBS{}{1}{}{}{1}{}{}{-1}
					\end{scope}	
					\begin{scope}[yshift=-4cm]
						\PBS{}{1}{}{}{1}{}{}{1}
					\end{scope}	
				\end{tikzpicture}
				}
			}
		\end{subfloatrow}
	}
	{
		\caption[Avoid list of figures error.]{
			\raggedright
			\begin{minipage}{\textwidth}
				\justify
				\small
				\vskip -3.4mm $\phantom{01234567}$ \textbf{The proposed experimental setup.}
				\subref{fig:1234} A Polarizing beam splitter (PBS) with two spatial input/output modes.
				\subref{fig:HH}-\subref{fig:VH} The effect of the PBS acting on the four possible two photon input states regarding polarizations.
				\subref{fig:LinFull} A post-selective linear optical scheme inducing a non-linear transformation \subref{fig:LinSimple}  and its simplified version utilising a polarizer reducing the success probability by $1/2$.
				\subref{fig:LinTwoScheme} A two level scheme amended by a unitary transformation $U$ acting on the polarization state of the photons. We consider a run of this experimental setup successful if all the detectors click. This condition introduces the post-selection to the system.				
			\end{minipage}
		}
		\label{fig:PBS}
	}
\end{figure}
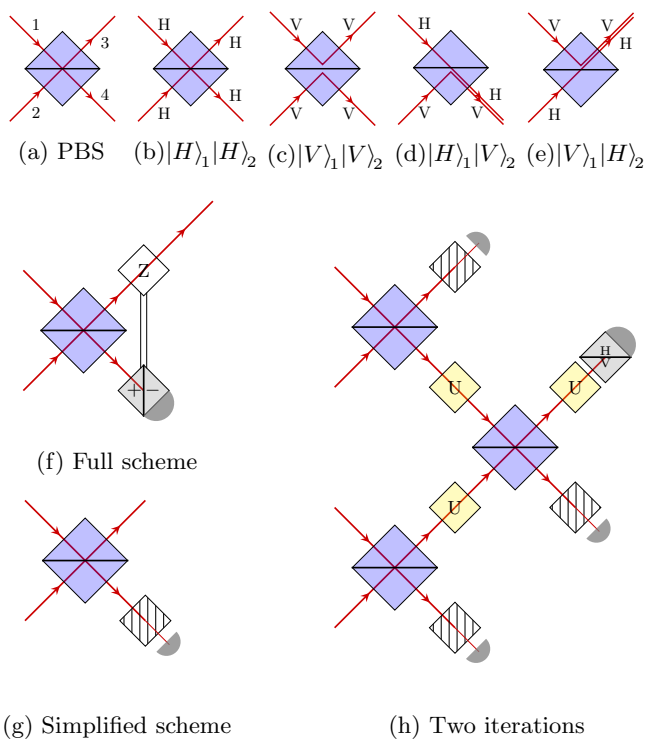

Let us denote the horizontal and vertical polarization states $\ket{H}$ and $\ket{V}$ for our photonic qubits. The key element of our scheme is the polarizing beam splitter (PBS). When two photons arrive at the same time but from different spatial input modes this linear optical element introduces entanglement between the spatial modes and the polarization degrees of freedom, see Fig.~\ref{fig:PBS}. We apply post-selection and accept the output of the PBS only if there is a photon in both spatial output modes.

Consider the effect of the PBS acting on a product state of two incoming photons:
\begin{equation*}
\begin{split}
(\alpha&\ket{H}+\beta\ket{V})_1\otimes(\alpha\ket{H}+\beta\ket{V})_2\rightarrow \\
&\alpha^2\ket{H}_3\ket{H}_4+\beta^2\ket{V}_3\ket{V}_4+
\alpha\beta(\ket{H,V}_{3}+\ket{H,V}_{4})
\end{split}
\end{equation*}
After post-selection the remaining quantum state is 
\begin{equation*}
\begin{split}
N&\!\left(\alpha^2\ket{H}_3\!\ket{H}_4\!+\!\beta^2\ket{V}_3\!\ket{V}_4\right)=\\
&\frac{N}{\sqrt{2}}\left((\alpha^2\ket{H}\!+\!\beta^2\ket{V})_3 
\ket{+}_4\!+\!(\alpha^2\ket{H}\!-\!\beta^2\ket{V})_3 
\ket{-}_4\right)
\end{split}
\end{equation*}
where $\ket{\pm}=(\ket{H}\pm\ket{V})/\sqrt{2}$ and $N=1/\sqrt{|\alpha|^4+|\beta|^4}$ is a norming factor. The success probability of the protocol is $1/N^2\geq 1/2$.
If we measure the photon at output mode $4$ the other photon collapses to $N\cdot\left(\alpha^2\ket{H}\pm\beta^2\ket{V}\right)$ corresponding to the measurement result. To get a definite outcome $\alpha^2\ket{H}+\beta^2\ket{V}$ we may apply a Pauli-Z gate whenever we measure $\ket{-}$ (Fig.~\ref{fig:LinFull}) or simply neglect such cases introducing another level of post-selection (Fig.~\ref{fig:LinSimple}). Either way the protocol implements a non-linear transformation $S: \alpha\ket{H}+\beta\ket{V} \mapsto N\cdot(\alpha^2\ket{H}+\beta^2\ket{V})$ which maps the identical qubit states of an ensemble to another qubit state of a smaller identical ensemble. If we iterate this process $S$ amended with an additional unitary step $U$, the iterates $(US)^n$ exhibit increasingly rich dynamics. 

\begin{figure}[ht]
	\phantomsection
	\centering
	\begin{subfigure}{.33\textwidth}
	  \centering
	  \includegraphics[width=0.91\textwidth]{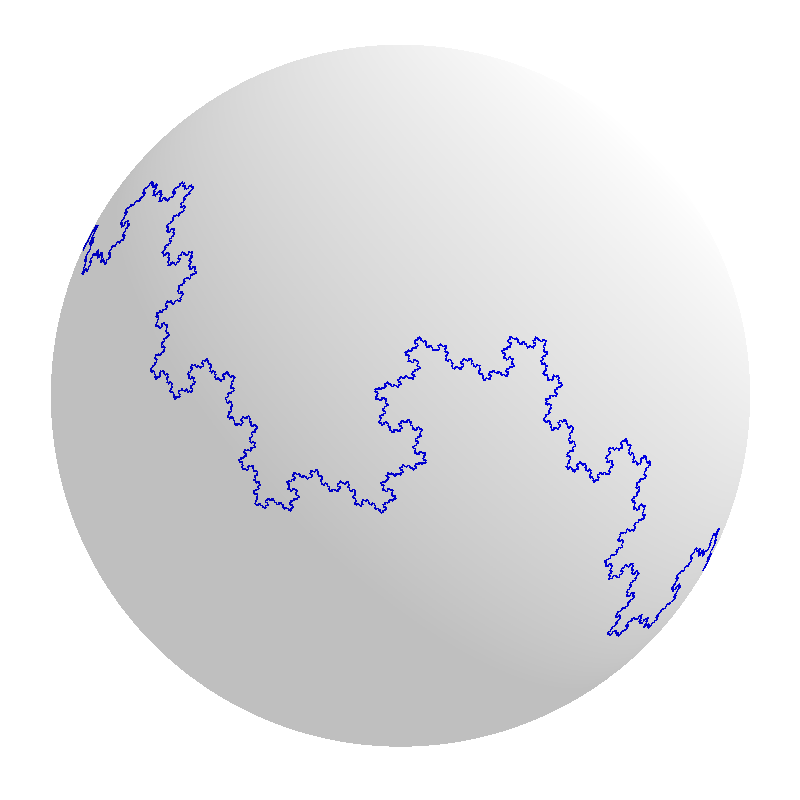}
	  \caption{$\theta=0.4,\varphi=\frac{\pi}{2}$}
	  \label{fig:aI}
	\end{subfigure}%
	\begin{subfigure}{.33\textwidth}
	  \centering
	  \includegraphics[width=0.91\textwidth]{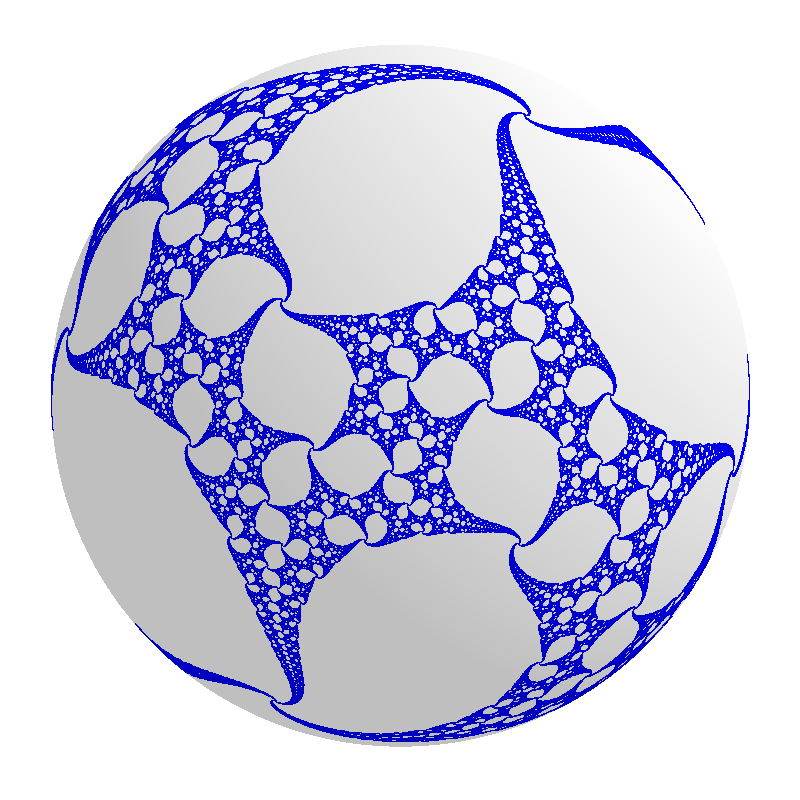}	  
	  \caption{$\theta=0.55,\varphi=\frac{\pi}{2}$}
	  \label{fig:bI}
	\end{subfigure}%
	\begin{subfigure}{.33\textwidth}
	  \centering
  	  \includegraphics[width=0.91\textwidth]{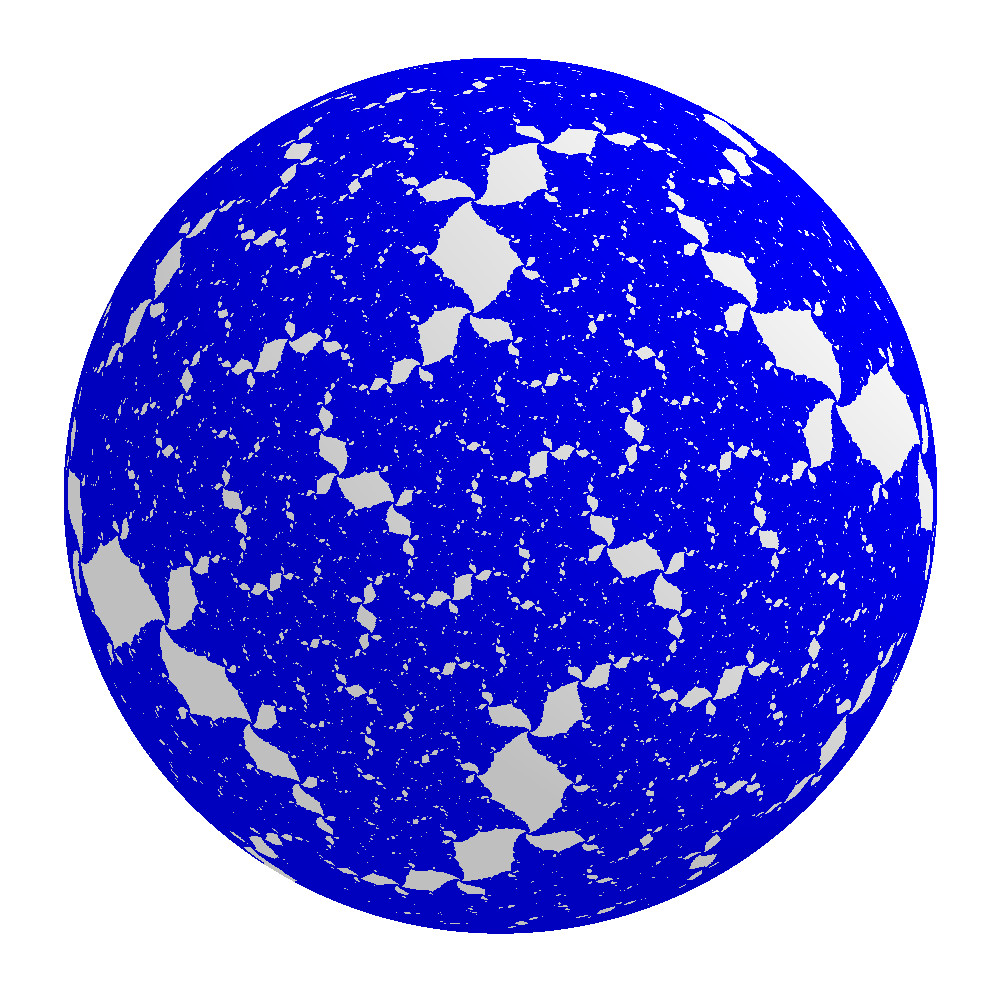}
	  \caption{$\theta=0.633,\varphi\!=\!\frac{\pi}{2}$}
	  \label{fig:cI}
	\end{subfigure}%
	\\
	\begin{subfigure}{.33\textwidth}
	  \centering
	  \includegraphics[width=0.91\textwidth]{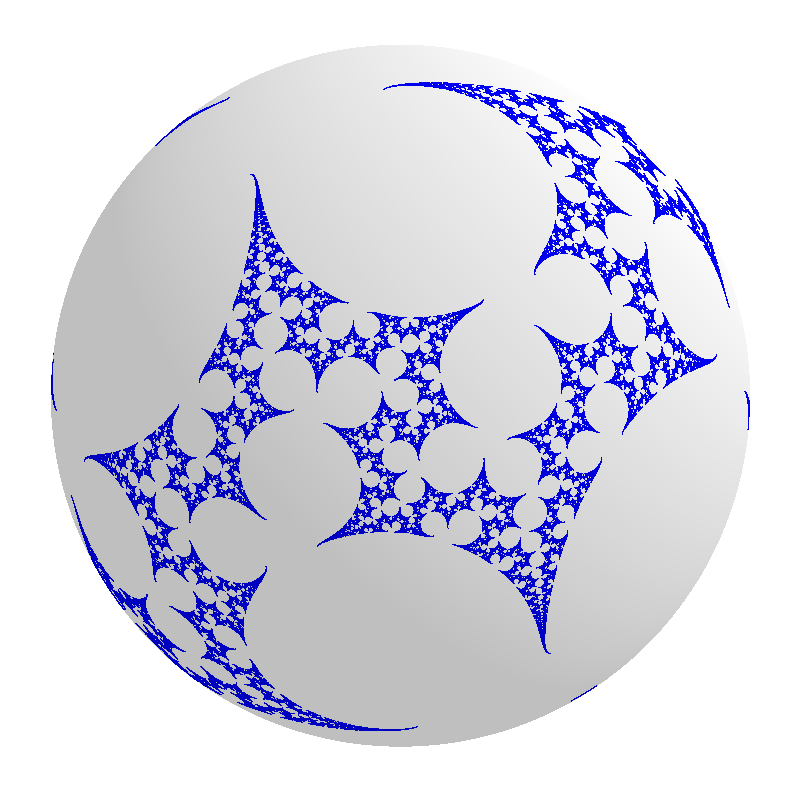}	 
	  \caption{$\theta=1.05,\varphi=\frac{\pi}{2}$}
	  \label{fig:dI}
	\end{subfigure}
	\begin{subfigure}{.33\textwidth}
	  \centering
	  \includegraphics[width=0.91\textwidth]{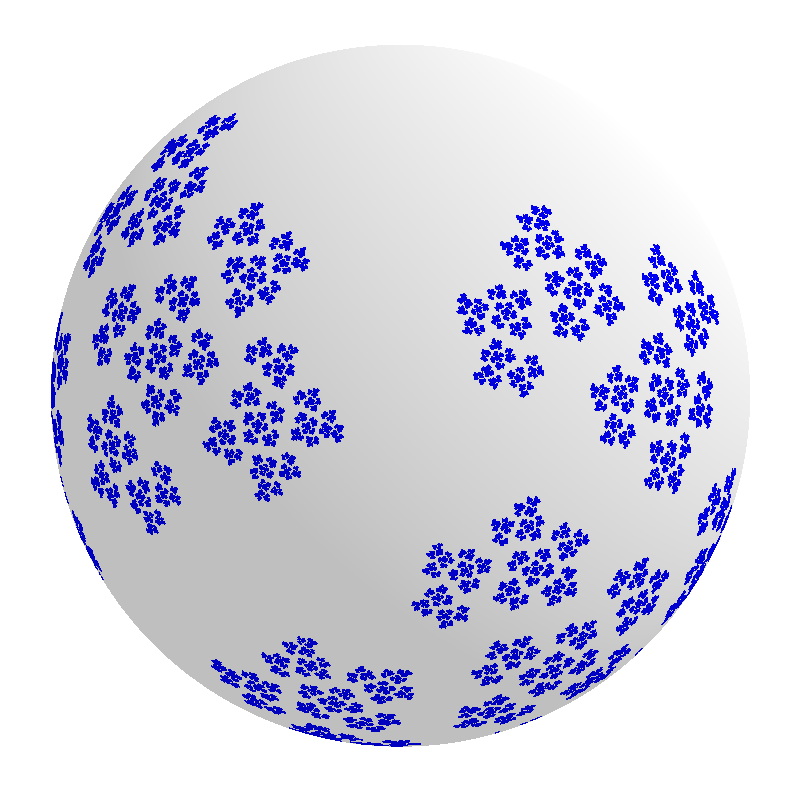}	 
	  \caption{$\theta\!=\!0.5, \varphi\!=\!0.5$}
	  \label{fig:bJ}
	\end{subfigure}%
	\begin{subfigure}{.32\textwidth}
	  \centering
	  \includegraphics[width=0.91\textwidth]{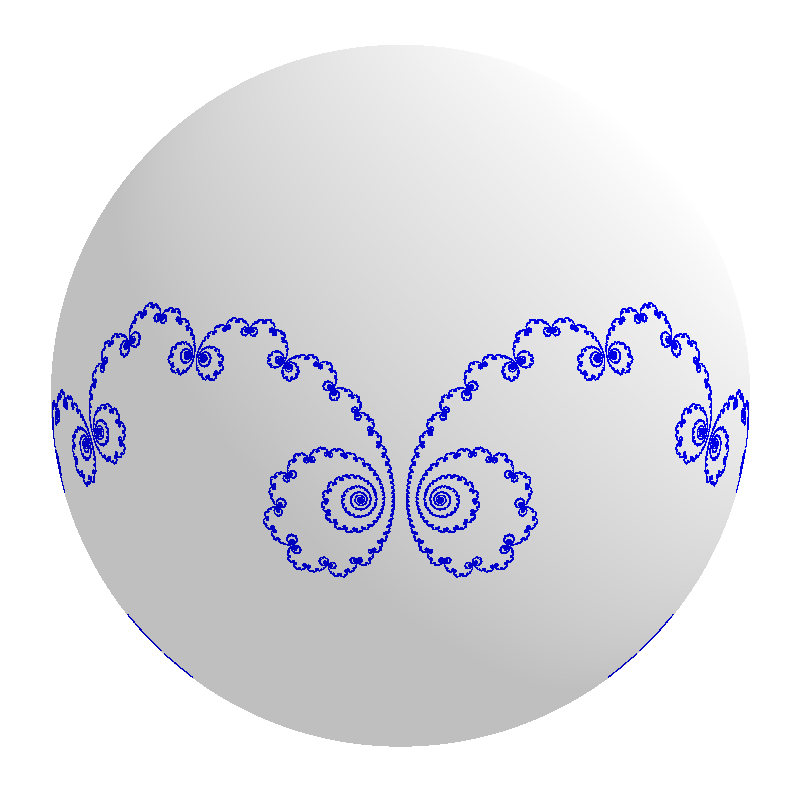}
	  \caption{$\theta\!=0.232, \varphi\!=0$}
	  \label{fig:dJ}
	\end{subfigure}
	\caption[Avoid list of figures error.]{
		\raggedright
		\begin{minipage}{\textwidth}
			\justify
			\small
			\vskip -3.2mm $\phantom{01234567}$
			\textbf{Julia sets consisting of the unstable initial states lying on the Bloch sphere. } Julia sets (blue) are plotted for various choices of $U_{\theta,\varphi}$.
			\subref{fig:aI}-\subref{fig:cI} Shows how the Julia set starts covering the whole Bloch sphere while increasing the value of $\theta$. \subref{fig:dI} Transition from a web to a simple closed curve. \subref{fig:bJ} The unstable states form a completely disconnected set. \subref{fig:dJ} The result of a parabolic explosion (implosion) \cite{Implosion} where a stable fixpoint has become unstable - breaking a circle like connected Julia set into infinitely many parts.	
		\end{minipage}
	}
	\label{fig:BlochJulia}
\end{figure}

It was shown \cite{OurPRA2006} that iteratively applying $US$ on an ensemble of identically prepared qubits the one qubit state of the ensemble $\ket{\psi_n}=\alpha_n\ket{H}+\beta_n\ket{V}$ after $n$ iterations may evolve sensitively with respect to the initial state $\ket{\psi_0}$. Nonetheless this was only shown for initial states lying on a zero-measure fractal on the Bloch sphere called the Julia set, see Fig.~\ref{fig:BlochJulia}. However certain choices of unitaries from the family
$U_{\theta,\varphi}=
	\left[\begin{array}{cc}
		\cos(\theta)	& \sin(\theta) e^{i\varphi}\\
		-\sin(\theta) e^{-i\varphi}	& \cos(\theta)
	\end{array}\right]$
seem to produce increasingly saturated Julia sets suggesting that it may reach a point where the whole Bloch sphere is covered by unstable initial states. A candidate for such a transformation is $\Phi=U_{\pi/4,\pi/2}S$ where
$U_{\pi/4,\pi/2}=\frac{1}{\sqrt{2}}\begin{bmatrix} 
1 & i \\ i & 1\\
\end{bmatrix}$. 

\vskip 4 mm \noindent \textbf{Exponentional sensitivity for all initial states.}
In order to handle the arising non-linear maps better we project the surface of the Bloch sphere to the complex plane using stereographic projection. Thus a (photonic) qubit $\ket{\psi}=\alpha\ket{H}+\beta\ket{V}$ may be described using a single complex parameter $z=\frac{\alpha}{\beta}\in\hat{\mathbb{C}}=\mathbb{C}\cup\{\infty\}$ including infinity. 
This representation yields ($\rightsquigarrow$) a new description of our protocol in terms of rational functions \nolinebreak\cite{MilnorBook,DevaneyBook}:
\begin{equation*}
\ket{\psi}\rightsquigarrow z;\,\,\, 
S \rightsquigarrow s: z \rightarrow z^2;\,\,\, 
\Phi \rightsquigarrow f: z \rightarrow \frac{z^2+i}{i z^2+1}
\end{equation*}

\newcommand{\scaleFig}{0.094}

\begin{figure}[ht]
	\phantomsection
	\centering
	\begin{subfigure}{.25\textwidth}
	  \centering
	  \includegraphics[scale=\scaleFig]{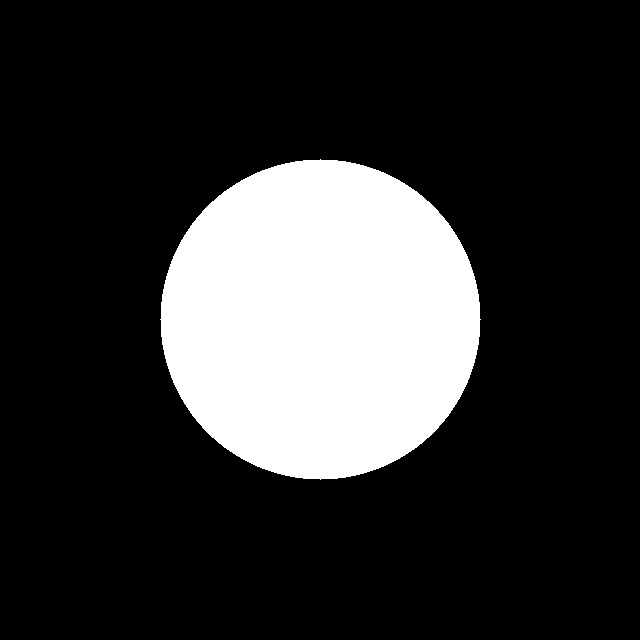}
	  \caption{$|z|\!>\!1$}
	  \label{fig:0}
	\end{subfigure}%
	\begin{subfigure}{.25\textwidth}
	  \centering
	  \includegraphics[scale=\scaleFig]{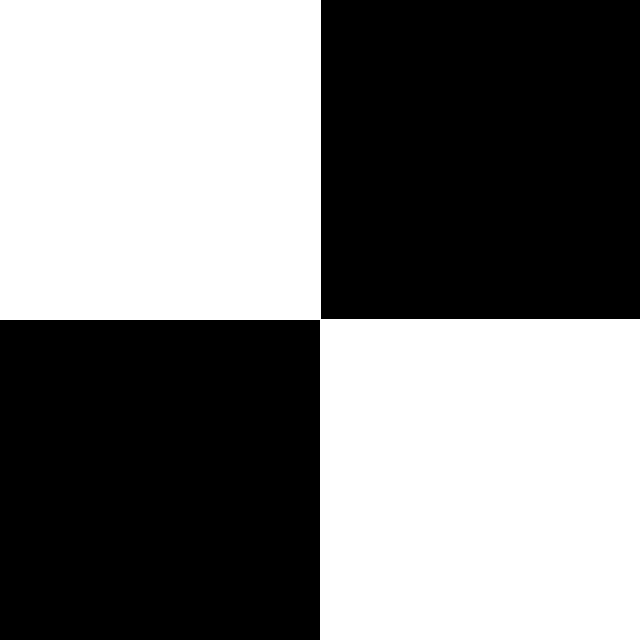}
	  \caption{$|f(z)|\!>\!1$}
	\end{subfigure}%
	\begin{subfigure}{.25\textwidth}
	  \centering
	  \includegraphics[scale=\scaleFig]{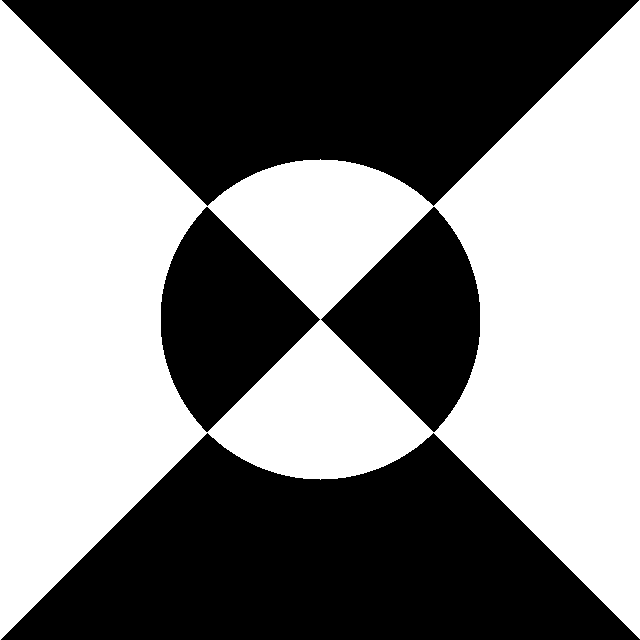}
	  \caption{$|f^{\circ 2}(z)|\!>\!1$}
	\end{subfigure}%
	\begin{subfigure}{.25\textwidth}
	  \centering
	  \includegraphics[scale=\scaleFig]{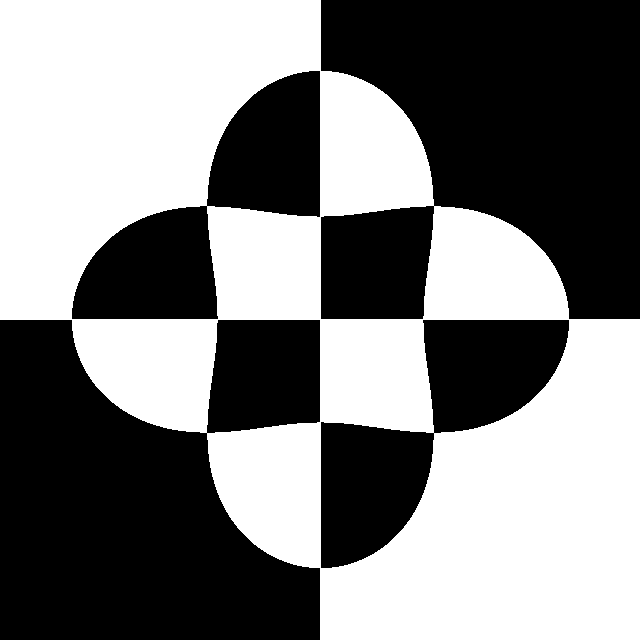}
	  \caption{$|f^{\circ 3}(z)|\!>\!1$}
	\end{subfigure}
	\\
	\begin{subfigure}{.25\textwidth}
	  \centering
	  \includegraphics[scale=\scaleFig]{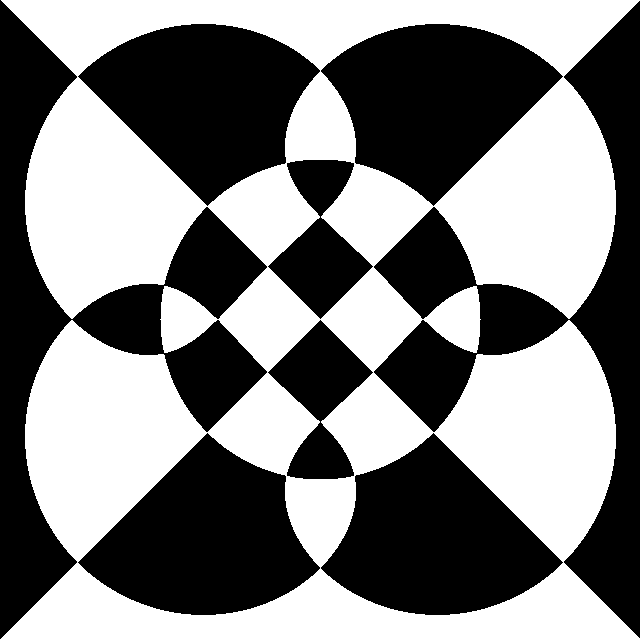}
	  \caption{$|f^{\circ 4}(z)|\!>\!1$}
	\end{subfigure}%
	\begin{subfigure}{.25\textwidth}
	  \centering
	  \includegraphics[scale=\scaleFig]{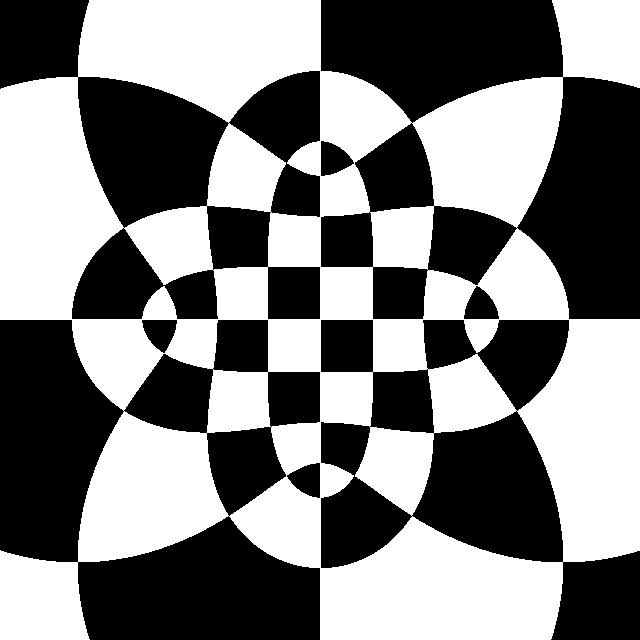}
	  \caption{$|f^{\circ 5}(z)|\!>\!1$}
	\end{subfigure}%
	\begin{subfigure}{.25\textwidth}
	  \centering
	  \includegraphics[scale=\scaleFig]{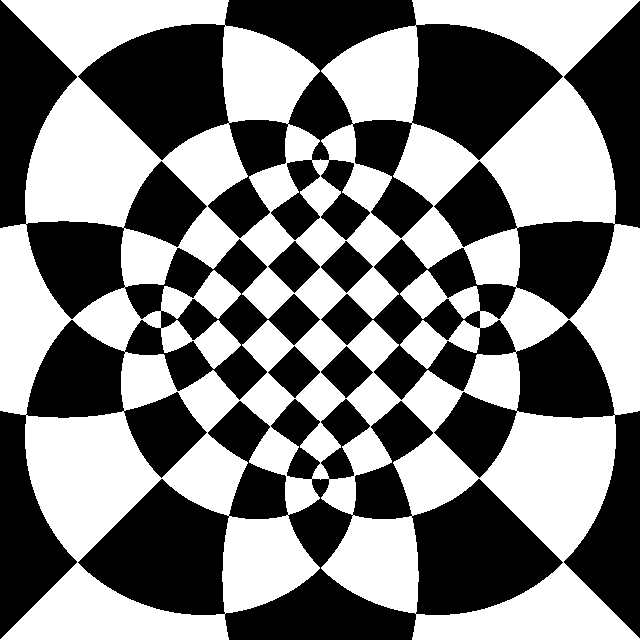}
	  \caption{$|f^{\circ 6}(z)|\!>\!1$}
	\end{subfigure}%
	\begin{subfigure}{.25\textwidth}
	  \centering
	  \includegraphics[scale=\scaleFig]{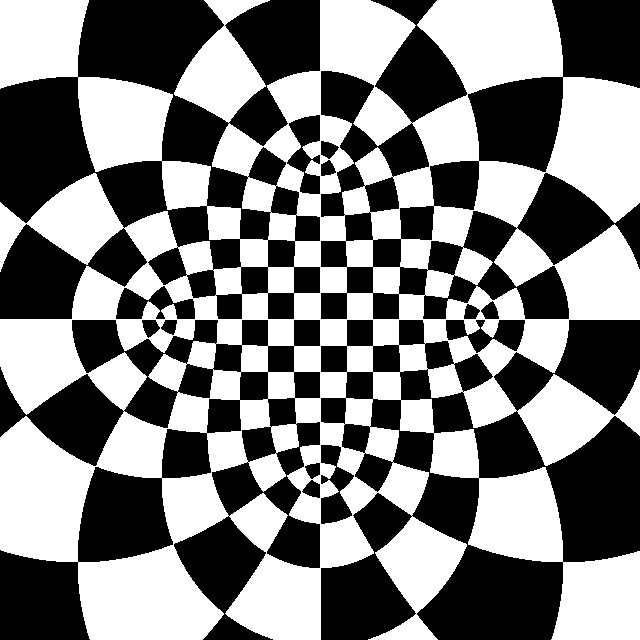}
	  \caption{$|f^{\circ 7}(z)|\!>\!1$}
	  \label{fig:7}
	\end{subfigure}
	\\
	\begin{subfigure}{.25\textwidth}
	  \centering
	  \includegraphics[scale=\scaleFig]{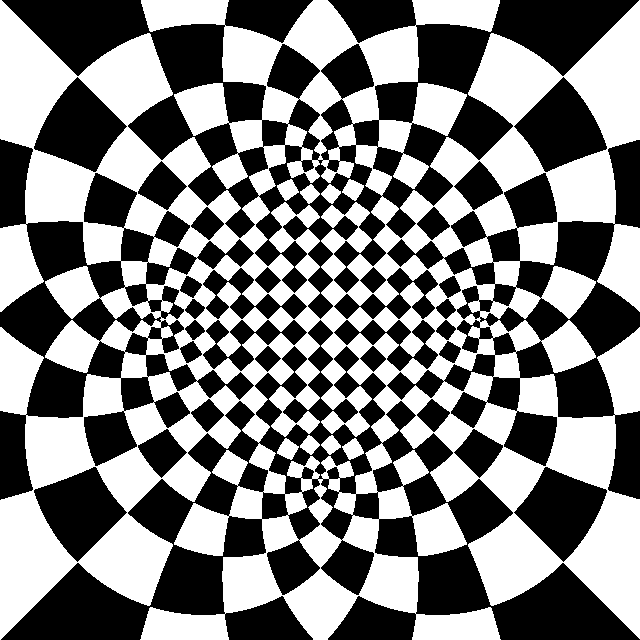}
	  \caption{$|f^{\circ 8}(z)|\!>\!1$}
	  \label{fig:8}
	\end{subfigure}%
	\begin{subfigure}{.25\textwidth}
	  \centering
	  \includegraphics[scale=\scaleFig]{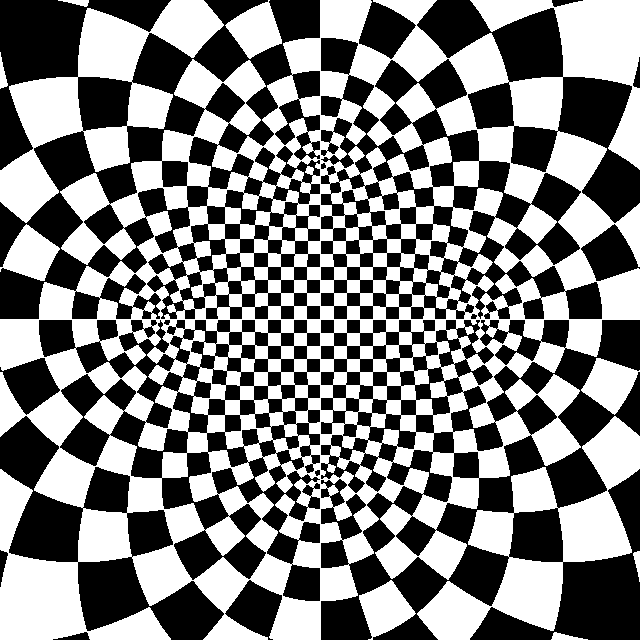}
	  \caption{$|f^{\circ 9}(z)|\!>\!1$}
	\end{subfigure}%
	\begin{subfigure}{.25\textwidth}
	  \centering
	  \includegraphics[scale=\scaleFig]{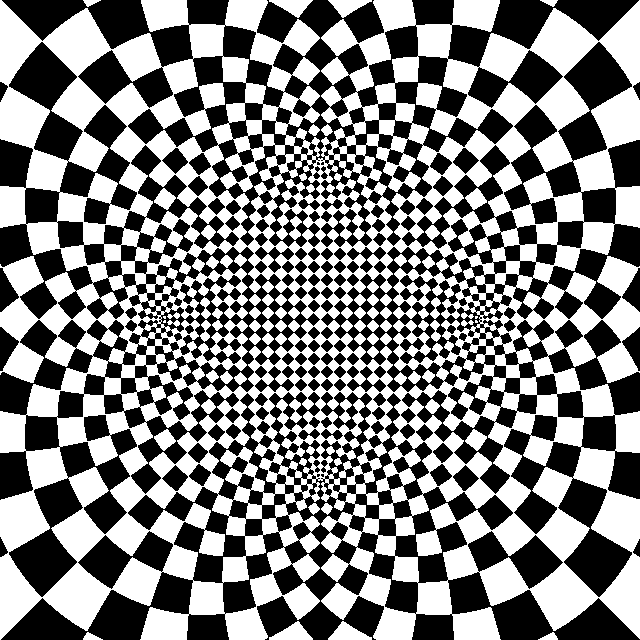}
	  \caption{$\!\!|f^{\circ 10}(z)|\!>\!1$}
	\end{subfigure}%
	\begin{subfigure}{.25\textwidth}
	  \centering
	  \includegraphics[scale=\scaleFig]{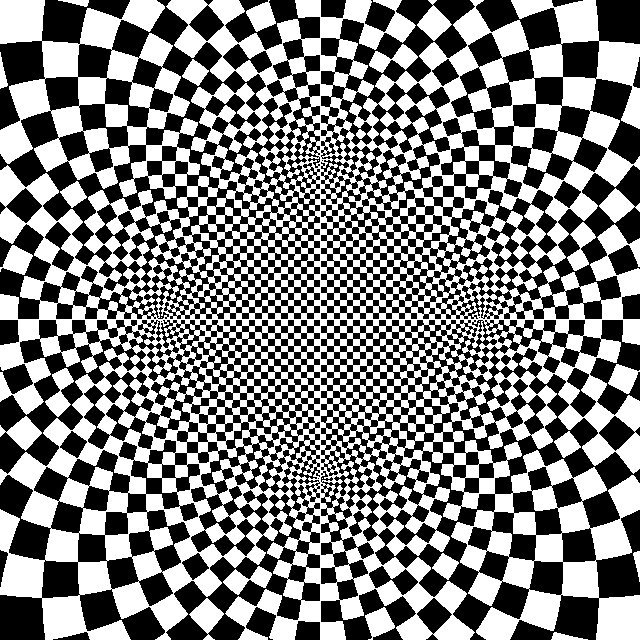}
	  \caption{$|f^{\circ 11}(z)|\!>\!1$}
  	  \label{fig:11}
	\end{subfigure}
	\\
	\vskip 0.6cm
	\begin{subfigure}{\textwidth}
	  \centering
		$\begin{tikzcd}
		\mathbb{C}/\mathbb{Z}[i]: \mqqquad \mqqquad \!\!& 
		\boxedtorus \arrow{r}{\!\!\!\cdot(1-i)} \arrow{d}{\wp} & 
		\boxedtorus \arrow{r}{\!\!\!\cdot(1-i)} \arrow{d}{\wp} &
		\cdots \arrow{r}{\!\!\!\cdot(1-i)} \arrow{d}{\wp} &
		\boxedtorus \arrow{d}{\wp} &\\
		\text{Bloch sphere}: \mqqquad
		& \sphere \arrow{r}{\Phi} & 
		\sphere \arrow{r}{\Phi} & 
		\overset{\phantom{.}}{\cdots\,} \arrow{r}{\Phi} & 
		\sphere & \\
		\end{tikzcd}$
		\vskip -1.0cm
		\caption{
			A commutative diagram explaining the action of $\Phi$.
		}
		\label{fig:cd}
	\end{subfigure}
	\caption[Avoid list of figures error.]{
		\raggedright
		\begin{minipage}{\textwidth}
			\justify
			\small
			\vskip -3.2mm $\phantom{01234567}$ \textbf{Iterations of an exponentially mixing map.}
			\subref{fig:0}-\subref{fig:11} Visualisation of the iteratives of $f$. Each subfigure shows $z\in [-2,2]\times[-2i,2i]$; the domains are coloured according to whether $|f_i^{\circ n}|>1$(black) or $\leq 1$ (white) distinguishing the northern and southern half of the Bloch sphere. 
			After a few iterations even very close states get mapped to different halves of the Bloch sphere as indicated by the rapid alternation of black and white domains.
			\subref{fig:cd} The iterative maps $\Phi^{\circ n}$ act on the Bloch sphere correspondingly to the action of multiplication by $(1-i)^n$ on the torus, explaining the regular pattern.
			\vskip -7 mm
			$\phantom{\sum}$
		\end{minipage}

	}
	\label{fig:chaosI}

\end{figure}

Using this formalism it turns out that $f$ is one of a few special so called Lattès maps~\cite{Lattes} and as such gained a lot of attention in the theory of complex dynamical systems~\cite{MilnorBook}. We can better understand the special properties of our Lattès map by analysing its relationship to the corresponding linear transformation of the $2$ dimensional torus. We will represent the torus $\mathbb{C}/\mathbb{Z}[i]$ as the complex plane modulo the Gaussian integers $\mathbb{Z}[i]=\{a+bi | a,b\in\mathbb{Z}\}$. Its transformation is represented by multiplication with $\cdot(1-i)$: $\tilde{z}\rightarrow (1-i)\cdot\tilde{z} \mod (1,i)$ which rotates and folds the torus $2$ times over itself. The correspondence between the torus and the sphere is established via the so called Weierstrass elliptic function $\wp: \mathbb{C}/\mathbb{Z}[i]\mapsto\hat{\mathbb{C}}$ \cite{HamiltonBook}. Relating the two surfaces gives rise to the identity $f^{\circ n}(z)=\wp((1-i)^n\cdot\wp^{-1}(z))$ showing that iterating $f$ on $\hat{\mathbb{C}}$ has essentially the same effect as repeatedly applying multiplication $\cdot(1-i)$ on $\mathbb{C}/\mathbb{Z}[i]$, see Fig.~\ref{fig:chaosI}. 

Viewing $\Phi$ through these glasses it becomes clear that it shows chaotic behaviour on the whole Bloch sphere. The map representing $\Phi$ on the torus uniformly stretches the surface of the torus by a factor of $\sqrt{2}$ and folds over itself two times. It is intuitively clear that the iterative application of such a transformation shows exponential sensitivity to the initial position on the torus and has a positive Lyapunov exponent. Even more strikingly it exhibits exponential mixing, yielding that close initial states separate exponentially fast on the surface of the Bloch sphere, as depicted on Fig.~\ref{fig:chaosI}. For a rigorous derivation of the exponential mixing see \ourappendix.

\pagebreak
\vskip 4 mm \noindent \textbf{Emergence of general complex rational dyanmics and the Mandelbrot set.}		
We may find even more exotic transformations by generalising the protocol allowing us the physical realisation of any rational map  $z \rightarrow \frac{\sum_{k=0}^n c_k z^k}{\sum_{k=0}^n d_k z^k} $ of degree $n\geq 2$. The generalised scheme proceed by forming $n$-tuples $\bigotimes_{j=1}^n(\alpha\ket{0}+\beta\ket{1})_j$ of identical pure qubits and applying an appropriate $n$ qubit unitary $V$. The final step is measuring all the qubits except the last one and keeping it only if all measurements resulted $0$; this post-selective step can be shortly described by the projection $\left[\ket{0}\!\!\bra{0\ldots00}+\ket{1}\!\!\bra{0\ldots01}\right]$. Implementing a specific rational map reduces to finding a $V$ unitary satisfying:
\begin{equation*}
	\begin{split}
		\left[\ket{0}\!\!\bra{0\ldots00}\right.\!&\left.+\ket{1}\!\!\bra{0\ldots01}\right]
		V\left(\otimes_{j=1}^n(\alpha\ket{0}+\beta\ket{1})_j\right) \\
		&= \gamma\left(\sum_{k=0}^n c_k \alpha^k\beta^{n-k}\ket{0}+\sum_{k=0}^n d_k \alpha^k\beta^{n-k}\ket{1}\right)
	\end{split}
\end{equation*}
with arbitrary $\gamma\neq 0$, providing the desired post selected state. The existence of such $V$ follows from a simple linear algebraic argument explicated in \ourappendix. 

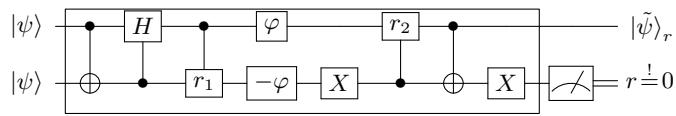
\begin{figure}[h]
	\phantomsection
	\centering
	\vskip 0.1cm
	\mbox{
		\Qcircuit @C=1em @R=1em {
		    \lstick{\ket{\psi}} & \ctrl{1}& \gate{H} & \ctrl{1}	& \gate{\varphi} &
				\qw	& \gate{r_2}	& \ctrl{1} & \qw & \qw &
				\rstick{\ket{\tilde{\psi}}_r} \qw\\
		    \lstick{\ket{\psi}} & \targ	& \ctrl{-1}	& \gate{r_1} & \gate{-\varphi} &	
			    \gate{X} & \ctrl{-1} & \targ & \gate{X}	& \meter &
		  		\rstick{\!r\!\underset{\phantom{0}}{\overset{!}{=}\!0}} \cw
			\gategroup{1}{2}{2}{9}{1.2em}{-}
		}
	}

	\caption[Avoid list of figures error.]{
		\raggedright
		\begin{minipage}{\textwidth}
			\justify
			\vskip -3.3mm $\phantom{01234567}$ \textbf{A quantum circuit implementing Mandelbrot maps.}
			 The controlled rotation of the gate labelled by $r_j$ is $\frac{1}{\sqrt{1+r_j^2}}\begin{bmatrix}1 & \phantom{-}r_j\\r_j & -1\end{bmatrix}$ and the phase gate labelled by $\varphi$ is $\begin{bmatrix}1 & 0\\0 & e^{i\varphi}\end{bmatrix}$ while H stands for a Hadamard gate. 
			If we set $r_1=1/r_2$, $r_2=|c|\cdot\sqrt{(1 + \sqrt{1 + 4/|c|^2})/2}$ and $e^{i\varphi}=\frac{c}{|c|}$ then the resulting map is $z^2+c$, if we accept only the $0$ measurement outcome on the second qubit.
		\end{minipage}
	}
	\label{fig:Mandel}
\end{figure}

A notable consequence is that we found a direct quantum physical realisation of the Mandelbrot maps $z\rightarrow z^2+c$ and can devise a quantum circuit for it. Fig.~\ref{fig:Mandel} shows a possible quantum circuit implementation for this family of maps. The scheme demonstrates how the corresponding family of $2$ qubit unitaries may be constructed using only controlled $1$ qubit gates which are considered experimentally more feasible in general. 

\vskip 4 mm \noindent \textbf{The cost of non-linearity and the Schrödinger microscope.}
We have just shown that using post-selection one can implement a wide range of non-linear maps that may be useful for various tasks. A highly non-linear map like we presented provides a sort of "Schrödinger microscope" \cite{SchrodingerMicroscope} enabling us to exponentially magnify tiny differences between quantum states. Having such a tool we are tempted to develop powerful quantum algorithms utilising it. It is well known that introducing post-selection to Quantum Computing makes it extremely powerful - the corresponding complexity class PostBQP~\cite{PostBQP} includes NP and even PP. The question of efficiency and resource needs naturally arises. We address it using a black box argument considering results about state discrimination \cite{QDet,Bergou,Herzog}.

Suppose we have a quantum device implementing $n$ iterations of $\Phi$ processing a qubit ensemble of size $N$. The size $M$ of the successfully processed output ensemble may be probabilistic. We would like to determine its success rate i.e. the average ratio $M/N$. To derive a bound on the success rate consider applying this quantum device to the qubit ensemble having state either $\ket{\psi_0}$ or $\ket{\psi_1}\neq \ket{\psi_0}$ with equal probability $1/2$. If their distance $d(\ket{\psi_0},\ket{\psi_1})=\sqrt{1-|\braket{\psi_0|\psi_1}|^2}\ll1/N$ then the distance of the full ensemble states is $d(\bigotimes^{N}\ket{\psi_0},\bigotimes^{N}\ket{\psi_1})\approx\sqrt{N}\cdot d(\ket{\psi_0},\ket{\psi_1})$ thus we cannot distinguish the possible ensemble states with error probability less than $\approx(1 -\sqrt{N}\cdot d(\ket{\psi_0},\ket{\psi_1}))/2$, see \cite[Chapter IV \S 2]{QDet}.
Suppose $d(\ket{\psi_j},\ket{+}) \ll1/2^n$ for $j=0,1$ then, after $n$ iterations of our process $\Phi$, the distance of the two sates increases by a factor of roughly $2^n$ since $\Phi$ doubles infinitesimal distances around the fixed point $\ket{+}$. 
If the device outputs $M$ copies of the transformed states then we can distinguish the ensembles with error probability $\approx(1 -\sqrt{M}\cdot2^n\cdot d(\ket{\psi_0},\ket{\psi_1}))/2$.
For large $N$ the success rate is roughly constant, because of the law of large numbers, thus we can treat the value $M/N$ fixed. But $\sqrt{M}\cdot2^{n}/\sqrt{N}$ cannot exceed $1$ as the error probability of discrimination cannot decrease and so the success rate is upper bounded by $4^{-n}$. This holds for states lying close to $\ket{+}$, in better cases the rate may be higher. For our implementation scheme each iteration has a success rate at least $1/4$ (up to a negligible term $-1/N$ due to parity) implying that this scheme provides the best possible worst case success rate.

In this way we showed that exponentially many copies are needed for $n$ iterations of the process. Similar upper bound can be devised to any non-linear map that have a region where the separation of close states can be described by a multiplicative factor $\lambda>1$. If we follow the above argument it turns out that the worst case success rate of such protocol is bounded by $1/\lambda^2$. Note that the particular choice of metric by which we measured separation is not limiting the scope of the argument too much -- we could use any other metric which agrees infinitesimally, e.g. the Bures metric. Thus it turns out that the implementation of any kind of Schrödinger microscope needs exponentially many copies of the states in terms of magnification steps, more precisely quadratically many in terms of the total magnification.

\vskip 4 mm \noindent \textbf{\large Discussion} \\* \noindent
While exploring the possible dynamical properties of state selective protocols we found that any complex rational map can be implemented using state selection. Such a general and natural correspondence between a physical system and the theory of complex dynamical systems is unique up to our knowledge. We could also devise a realistic optical experimental scheme which implements particularly interesting quadratic rational dynamics.

Meanwhile we studied the emergence of exponential sensitivity and it lead us to the analysis of implementation cost which turned out to be exponential. We found a general bound on the number of copies needed for the successful implementation of any expanding non-linear map. We proved that a protocol capable of magnifying differences between close quantum states by a factor $\lambda>1$ necessarily yields a rate of loss at least $1-1/\lambda^2$ in the number of copies of the unknown input quantum states. This "Quantum magnification bound" is basically another reformulation of the fact that one cannot bootstrap quantum information without an external source and somewhat resembles the quantum no-cloning theorem.

We used the "Quantum magnification bound" principle to show the optimality of our implementation of an exponentially mixing map. Similarly, it may be applied to other quantum information protocols providing a general tool for bounding the success rate of particular probabilistic protocols. This principle also helps to understand the advances and limitations of any kind of Schrödinger microscope regardless the actual implementation method. The term Schrödinger microscope was first introduced for a protocol using collective weak measurements and coherent feedback \cite{SchrodingerMicroscope} where the limitations were unclear due to the approximate arguments applied. A fundamental approach like the one we presented should be helpful for understanding such complex systems as well.

Looking at general processes with inspiration coming from this principle may also provide a new insight to the relation of classical and quantum chaos \cite{MesoNano}, suggesting that classical deterministic chaos may be just an approximation with a characteristic time scale. Classical deterministic chaotic systems explode fast and observing the deterministic evolution of the system even at a macroscopic level enables the determination of the initial conditions increasingly precisely \cite{InfoGain}. But there is a level of precision that is prohibited by quantum uncertainty relations. This is an apparent philosophical contradiction if we believe classical physics is based on quantum mechanics; a possible dissolution is saying that on long time scales one cannot treat a classical process deterministically chaotic just chaotic in some statistical sense.

\vskip 4 mm \noindent \textbf{\large Acknowledgements} \\* \noindent
We thank Dr. János Asbóth and Dr. András Frigyik for stimulating discussions.
We acknowledge support by GACR 13-33906S, RVO 68407700, the Hungarian Scientific Research Fund (OTKA) under Contract Nos. K83858, NN109651, the Hungarian Academy of Sciences (Lendület Program, LP2011-016) and the Deutscher Akademischer Austauschdienst  (DAAD  project  no.  65049).



\begin{thebibliography}{1000}

\bibitem{QControl} H. Rabitz, \href{http://dx.doi.org/10.1088/1367-2630/11/10/105030}{Focus on Quantum Control, New J. Phys. 11, 105030 (2009)}

\bibitem{KLM} E. Knill, R. Laflamme, G. J. Milburn: 
\href{http://dx.doi.org/10.1038/35051009}{A scheme for efficient quantum computation with linear optics, Nature 409, 46–52 (2001)}

\bibitem{Calsamiglia} B. Gendra, E. Ronco-Bonvehi, J. Calsamiglia, R. Muñoz-Tapia, E. Bagan: 
\href{http://dx.doi.org/10.1103/PhysRevLett.110.100501}{Quantum Metrology Assisted by Abstention, Phys. Rev. Lett. 110, 100501 (2013)}

\bibitem{QRepeater} L.M. Duan, M.D. Lukin, J.I. Cirac, and P. Zoller,
\href{http://dx.doi.org/10.1038/35106500}{Long-distance quantum communication with atomic ensembles and linear optics, Nature 414, 413-418 (2001)}

\bibitem{purification1} C. H. Bennett, G. Brassard, S. Popescu, B. Schumacher, J.A. Smolin, and W.K. Wootters, \href{http://dx.doi.org/10.1103/PhysRevLett.76.722}{Purification of Noisy Entanglement and Faithful Teleportation via Noisy Channels, Phys. Rev. Lett. 76, 722 (1996)}

\bibitem{purification2} C.H. Bennett, D.P. DiVincenzo, J.A. Smolin, and W.K. Wootters, \href{http://dx.doi.org/10.1103/PhysRevA.54.3824}{Mixed-state entanglement and quantum error correction, Phys. Rev. A 54, 3824 (1996)}

\bibitem{purification3} D. Deutsch, A. Ekert, R. Jozsa, C. Macchiavello, S. Popescu, and A. Sanpera, \href{http://dx.doi.org/10.1103/PhysRevLett.77.2818}{Quantum Privacy Amplification and the Security of Quantum Cryptography over Noisy Channels, Phys. Rev. Lett. 77, 2818 (1996)}

\bibitem{Gisin} H. Bechmann-Pasquinucci, B. Huttner, and N. Gisin,
\href{http://dx.doi.org/10.1016/S0375-9601(98)00189-3}{Non-linear quantum state transformation of spin-1/2. Phys.Lett. A242, 198-204, (1998)} 

\bibitem{Scott} A. J. Scott, G. J. Milburn, \href{http://dx.doi.org/10.1103/PhysRevA.63.042101}{Quantum nonlinear dynamics of continuously measured systems, Phys. Rev. A 63, 042101 (2001)}

\bibitem{Habib} S. Habib, K. Jacobs, K. Shizume, \href{http://dx.doi.org/10.1103/PhysRevLett.96.010403}{Emergence of Chaos in Quantum Systems Far from the Classical Limit, Phys. Rev. Lett. 96, 010403 (2006)}

\bibitem{Everitt} M. J. Everitt, \href{http://dx.doi.org/10.1088/1367-2630/11/1/013014}{On the correspondence principle: implications from a study of the nonlinear dynamics of a macroscopic quantum device }

\bibitem{ChaosBook} P. Cvitanovi\'c, R. Artuso, R. Mainieri, G. Tanner and G. Vattay,
\href{http://chaosbook.org/}{Chaos: Classical and Quantum, {\tt chaosbook.org}, Niels Bohr Institute, Copenhagen (2012)}

\bibitem{OurPRA2006} T. Kiss, I. Jex, G. Alber, and S. Vym\v{e}tal, \href{http://dx.doi.org/10.1103/PhysRevA.74.040301}{Complex chaos in the conditional dynamics of qubits, Phys. Rev. A 74, 040301(R) (2006)}

\bibitem{2011} T. Kiss, S. Vymetal, L.D. Toth, A. Gabris, I. Jex, and G. Alber, \href{http://dx.doi.org/10.1103/PhysRevLett.107.100501}{Measurement induced chaos with entangled states, Phys. Rev. Lett. 107, 100501 (2011)}

\bibitem{Guan} Y. Guan, D. Q. Nguyen, J. Xu, and J. Gong, \href{http://dx.doi.org/10.1103/PhysRevA.87.052316} {Reexamination of measurement-induced chaos in entanglement-purification protocols. Phys. Rev. A, 87, 052316 (2013)} 

\bibitem{DevaneyBook} R.L. Devaney, 
\href{http://math.bu.edu/people/bob/books.html}{{\it An Introduction to Chaotic Dynamical Systems}, Westview Press (2003)}

\bibitem{PBS} J.W. Pan, C. Simon, S. Brukner \& A. Zeilinger: 
\href{http://dx.doi.org/10.1038/35074041}{Entanglement purification for quantum communication, Nature 410, 1067-1070 (2001)}

\bibitem{PBSExperiment} J.W. Pan, S. Gasparoni, R. Ursin, G. Weihs \& A. Zeilinger: 
\href{http://dx.doi.org/10.1038/nature01623}{Experimental entanglement purification of arbitrary unknown states, Nature 423, 417-422 (2003)}

\bibitem{MilnorBook} J. Milnor, \href{http://press.princeton.edu/titles/8117.html}{{\it Dynamics in One Complex Variable}, Princeton University Press (2006)}


\bibitem{Implosion} A. Douady,
\href{http://dx.doi.org/10.1090/psapm/049/1315535}{{\it Does a Julia set depend continuously on the polynomial? in Complex dynamical systems: The mathematics behind the Mandelbrot set and Julia sets (R. L. Devaney, ed.)},  Proc. of Symposia in Applied Math. vol. 49, pp. 91–138., Amer. Math. Soc. (1994)}

\bibitem{Lattes} J. Milnor, \href{http://arxiv.org/abs/math/0402147}{On Lattès maps. In Dynamics on the Riemann Sphere, a Bodil Branner Festschrift, P. Hjorth and C. L. Petersen, editors, pp. 9–43. Eur. Math. Soc. Zürich (2005)}

\bibitem{HamiltonBook} A.V. Bolsinov and A.T. Fomenko, \href{http://www.crcpress.com/product/isbn/9780415298056}{{\it Integrable Hamiltonian Systems: Geometry, Topology, Classification}, CRC Press (2004)}

\bibitem{SchrodingerMicroscope} S. Lloyd and J.E. Slotine, \href{http://dx.doi.org/10.1103/PhysRevA.62.012307}{Quantum feedback with weak measurements, Phys. Rev. A 62, 012307 (2000)}

\bibitem{PostBQP} S. Aaronson, \href{http://arxiv.org/abs/quant-ph/0412187}{Quantum Computing, Postselection, and Probabilistic Polynomial-Time, arXiv:quant-ph/0412187 (2004)}

\bibitem{QDet} Carl W. Helstrom, \href{http://books.google.es/books/about/Quantum_Detection_and_Estimation_Theory.html?hl=en&id=fv9SAAAAMAAJ}{{\it Quantum detection and estimation theory}, Academic Press (1976)}

\bibitem{Bergou} E. Bagan, R. Muñoz-Tapia, G. A. Olivares-Rentería, J. A. Bergou, \href{http://dx.doi.org/10.1103/PhysRevA.86.040303}{Optimal discrimination of quantum states with a fixed rate of inconclusive outcomes, Phys. Rev. A 86, 040303(R) (2012)}

\bibitem{Herzog} U. Herzog, \href{http://dx.doi.org/10.1103/PhysRevA.86.032314}{Optimal state discrimination with a fixed rate of inconclusive results: Analytical solutions and relation to state discrimination with a fixed error rate, Phys. Rev. A 86, 032314 (2012)}

\bibitem{MesoNano} M. G. E. da Luz, C. Anteneodo, \href{http://dx.doi.org/10.1098/rsta.2010.0301 }{Nonlinear dynamics in meso and nano scales: fundamental aspects and applications, Phil. Trans. Roy. Soc. A vol. 369, pp. 245-259, i. 1935 (2011)}

\bibitem{InfoGain} V. Madhok, C.A. Riofrío, S. Ghose, I.H. Deutsch, \href{http://dx.doi.org/10.1103/PhysRevLett.112.014102}{Information Gain in Tomography–A Quantum Signature of Chaos, Phys. Rev. Lett. 112, 014102 (2014)}



\end{thebibliography}

\vskip 4 mm \noindent \textbf{\large \ourappendix} \\* \noindent
\textbf{Connection to Lattès maps.} 
The map $f=\frac{z^2+i}{i z^2+1}$ can be written as $\frac{z+1}{z-1}\circ\frac{z^2+1}{-2iz}\circ\frac{z+1}{z-1}$, where $\tilde{f}=\frac{z^2+1}{-2iz}$ is a well-known Lattès map~\cite{Lattes}, and $\frac{z+1}{z-1}$ is a self inverse Möbius transformation~\cite{MilnorBook}. Since conjugation by Möbius transformation do not change the iterative features it implies that $f$ exhibits the same dynamics as $\tilde{f}$. To give a physical meaning to this Möbius transformation we mention that it corresponds to a rotation of the Bloch sphere i.e. $f$ and $\tilde{f}$ essentially describe the same process just written in a different qubits basis. 

As we already indicated in this article $f$ is conjugate to the map $(1-i)\cdot\text{id}_{\mathbb{C}/\mathbb{Z}[i]}$ via the Weierstrass-$\wp$ function. In fact, we need a slightly transformed version of the Weierstrass-$\wp$ which amended by our Möbius transformation. Our transformed version can be written as $\tilde{\wp}(z):=\frac{z+1}{z-1}\circ \frac{2\wp(z;L =i)}{i\sqrt{g_2(L=i)}}$ (for the definitions of $ L, g_2$ and $\wp(z; L)$ see \cite{HamiltonBook}). Using this function we get the key identity $f^{\circ n}\circ\tilde{\wp}=\tilde{\wp}\circ(1-i)^n \cdot\text{id}_{\mathbb{C}/\mathbb{Z}[i]}$. 

$\tilde{\wp}$ induces a two sheet branched covering with $4$ exceptional points which are covered only once. (These points can be easily spotted on Fig.~\ref{fig:chaosI}). With the exception of these $4$ points $\tilde{\wp}$ is a two-to-one map so it does not have a well defined inverse. A general point $\hat{z}\in \mathbb{C}$ has pre-images $z,-z \in \mathbb{C}/\mathbb{Z}[i]$, but the linear map on $\mathbb{C}/\mathbb{Z}[i]$ carries opposite numbers to opposite ones so fortunately the identity 
$f(\hat{z})=\tilde{\wp}((1-i)\cdot\tilde{\wp}^{-1}(\hat{z}))\,\, \forall \hat{z}\in \hat{\mathbb{C}}$ holds regardless which branch of $\tilde{\wp}^{-1}$ is considered. In this sense we can say that the stronger identity  
$f^{\circ n}(\hat{z})=\tilde{\wp}((1-i)^n\cdot\tilde{\wp}^{-1}(\hat{z}))\,\, \forall \hat{z}\in \hat{\mathbb{C}}$ also holds. 

\vskip 4 mm \noindent \textbf{Metric on the Bloch sphere $\hat{\mathbb{C}}$ and the torus $\mathbb{C}/\mathbb{Z}[i]$.}
We would like to show exponential mixing of our Lattès map thus we need to understand how distances are distorted by $\tilde{\wp}$. In order to trace the problem we need to introduce some proper distance concepts.

A possible metric on pure quantum states is given by using the distance defined by the quantum angle $d_A(\ket{\psi_{z_0}},\ket{\psi_z}):=\arccos{|\braket{\psi_0|\psi_z}|}$. Note that this distance coincides with the natural spherical metric of the Bloch sphere up to a multiplicative factor of $2$. This metric is similar to the Bures metric defined by the distance
$d_B=\sqrt{2(1-F(\ket{\psi_{z_0}}\!\bra{\psi_{z_0}},\ket{\psi_{z}}\!\bra{\psi_{z}}))}$, where $F(.,.)$ is the Fidelity of two density matrices. A third possible distance definition is given by $d(\ket{\psi_{z_0}},\ket{\psi_z})=\sqrt{1-|\braket{\psi_{z_0}|\psi_{z}}|^2}$. Since all three distance definitions coincide for infinitesimal distances we are free to chose the most appropriate one for our calculations. For now we stick with the natural metric of $\hat{ \mathbb{C}}$ which is 2 times the quantum angle $d_R(z_0,z)=2\cdot d_A(\ket{\psi_{z_0}},\ket{\psi_z})$, the index $R$ refers to the fact that $d_R$ is a Riemannian metric. For the torus $\mathbb{C}/\mathbb{Z}[i]$ we use the natural Riemannian metric inherited from $\mathbb{C}$. 

First we show that the Möbius transformation $\frac{z+1}{z-1}$ leaves the metric on $\hat{\mathbb{C}}$ invariant. I.e. we would like to show that the conformal metric $\rho:=\frac{|ds'|}{|ds|}$ is trivial, where $ds$ and $ds'$ are tangent vectors of $\hat{\mathbb{C}}$ at points $s$ and $s'=\frac{s+1}{s-1}$ such that $ds$ is mapped to $ds'$ by the tangent map. We proceed using the identity $\frac{|ds'|}{|ds|}=\frac{|dz|}{|ds|}\frac{|dz'|}{|dz|}\frac{|ds'|}{|dz'|}$, where $dz$ and $dz'$ are tangent vectors of $\mathbb{C}$. The conformal metric transformation introduced by the stereographic projection is well known to be $\frac{|ds'|}{|dz'|}=\frac{4}{|z'|^2+1}$, similarly $\frac{|dz|}{|ds|}=\frac{|z|^2+1}{4}$. 
Finally $\frac{|dz'|}{|dz|}=\left|d\left(\frac{z+1}{z-1}\right)/dz \right|=\left|\frac{(z-1)-(z+1)}{(z-1)^2}\right|=\frac{2}{|z-1|^2}$. Putting everything together $\frac{|ds'|}{|ds|}=\frac{|z|^2+1}{4}\frac{2}{|z-1|^2}\frac{4}{|\frac{z+1}{z-1}|^2+1}=\frac{|z|^2+1}{1}\frac{2}{|z-1|^2}\frac{|z-1|^2}{|z+1|^2+|z-1|^2}=1$ as we indicated.

The Möbius transformation is just an isometry of $\hat{\mathbb{C}}$ so we can concentrate on the other part $\frac{2\wp(z;L =i)}{i\sqrt{g_2(L=i)}}$ of our Weierstrass function $\tilde{\wp}$. Now $\rho:=\frac{|ds|}{|dz|}=\frac{|dz'|}{|dz|}\frac{|ds|}{|dz'|}$ where $dz'$ is a tangent vector of $\mathbb{C}$ at a point $z'$. Just as above $\frac{|ds|}{|dz'|}=\frac{4}{|z'|^2+1}=\frac{4}{\frac{4}{g_2}|\wp(z)|^2+1}$. For the other factor $\frac{|dz'|}{|dz|}=|d\frac{2\wp(z;L =i)}{i\sqrt{g_2(L=i)}}/dz|= \frac{2}{\sqrt{g_2}}|\wp'(z)|$. Now we use the well known property ${\wp'}^2=4\wp^3-g_2\wp-g_3$. Since $g_3(L=i)=0$ the final formula is:
$\rho^2=\frac{16}{(\frac{4}{g_2}|\wp(z)|^2+1)^2}\frac{4}{g_2}|4\wp(z)^3-g_2\wp(z)|= 64g_2\frac{|4\wp(z)^3-g_2\wp(z)|}{(4|\wp(z)|^2+g_2)^2}$. Using the triangle inequality we get $\rho^2\leq 64g_2\frac{|4\wp(z)^3|+|g_2\wp(z)|}{(4|\wp(z)|^2+g_2)^2}= 64g_2\frac{|\wp(z)|}{4|\wp(z)|^2+g_2}$. This function has its maximum when $|\wp(z)|^2=g_2/4$, substitution yields $\rho^2\leq16\sqrt{g_2}$. Finally using that $\sqrt{g_2}\approx 13.7504$ we arrive at the conclusion $\rho<16$. 

The conformal metric $\rho<16$ is upper bounded meaning that the image of any two points from the surface of the torus gets mapped to points having spherical distance less than $16$ times their torical distance. Looking at this inequality the other way round shows that if have a point $s$ on the surface of the sphere and a radius $\varepsilon$ ball $B_{d_R}(s,\varepsilon)$ around it then this ball's pre-image $\tilde{\wp}^{-1}(B_{d_R}(s,\varepsilon))$ contains another ball $B_{d_{\mathbb{C}/\mathbb{Z}[i]}}(\tilde{\wp}^{-1}(s),\varepsilon/16)$.

\vskip 4 mm \noindent \textbf{Exponential mixing.}
Suppose we have a one qubit state $\ket{\psi_{z}}$ prepared with accuracy $\varepsilon$ meaning that it is a pure quantum state $\ket{\psi_z}$ close to some $\ket{\psi_{z_0}}$ such that the quantum angle $d_A(\ket{\psi_{z_0}},\ket{\psi_z})=\arccos{|\braket{\psi_{z_0}|\psi_z}|}\leq \varepsilon$. In other words $\ket{\psi_{z}}$ lies in the diameter $\varepsilon$ ball around $\ket{\psi_{z_0}}$ i.e. $\ket{\psi_{z}}\in B_{d_A}(\ket{\psi_{z_0}},\varepsilon)$. This ball corresponds to $B_{d_R}(z_0,2\varepsilon)$ using the (Bloch) spherical representation. As we already discussed
$\tilde{\wp}^{-1}(B_{d_R}(s,2\varepsilon))\supset B_{d_{\mathbb{C}/\mathbb{Z}[i]}}(\tilde{\wp}^{-1}(s),\varepsilon/8)$.
 
It is easy to show that after $n=-\log_{\sqrt{2}}(\varepsilon/8)$ iteration of $\cdot(1-i)$ over all the points of $B_{d_{\mathbb{C}/\mathbb{Z}[i]}}(\wp^{-1}(z_0),\delta)$ covers the whole $\mathbb{C}/\mathbb{Z}[i]$. Thus for $n=6-\log_{\sqrt{2}}(\varepsilon)$ we have $\hat{\mathbb{C}}=f^{\circ n}(B_{d_A}(z_0,\varepsilon))$. It means that after $6-\log_{\sqrt{2}}(\varepsilon)$ iterations the initial uncertainty about the state $\ket{\psi_z}$ evolves so much that $\Phi^{\circ n}(\ket{\psi_z})$ may be any pure state. This statement is basically a translation of the fact that the linear map $\cdot(1-i)$ has Lyapunov exponent $\ln(\sqrt{2})$ on $\mathbb{C}/\mathbb{Z}[i]$ and shows exponential mixing and sensitivity on the surface of the Bloch sphere.

\vskip 4 mm \noindent \textbf{Construction of $n$-qubit unitaries for degree $n$ rational maps.} 
We would like to implement the rational function $z \rightarrow \frac{\sum_{k=0}^n c_k z^k}{\sum_{k=0}^n d_k z^k} $. Our generalised protocol starts by forming $n$-tuples of identical pure qubits of our ensemble then continues by the application of a specific $n$ qubit unitary $V$. The final step is a measurement on all the qubits except the last one of every tuple. The protocol succeeds if all the measurements resulted in $0$, the unmeasured qubit is kept only in such cases. 

Initially the state of the $n$-tuples is the following product state: 
\begin{equation*}
\begin{split}
\ket{\phi}=\bigotimes_{i=1}^n(\alpha\ket{0}_i+\beta\ket{1}_i)=  \sum_{k=0}^n \alpha^k\beta^{n-k}\ket{\phi_k}\\
\text{ where }\ket{\phi_k}=\sum_{\underset{\#\{b_i| b_i=1\}=k}{b\in\{0,1\}^n}}\ket{b}
\end{split}
\end{equation*}

As before we use the parametrisation $z=\alpha/\beta$ for a qubit $(\alpha\ket{0}+\beta\ket{1})$. Then the parameter of the unmeasured, post selected qubit can be described as follows:

\begin{equation}
\begin{split}
\label{eq:nmap}
\frac{\braket{\overset{n-1 \text{ zeros}}{\overbrace{0\ldots0}}0|V|\phi}}{\braket{\underset{n-1 \text{ zeros}}{\underbrace{0\ldots0}}1|V|\phi}}
&=\frac{\sum_{k=0}^n\overset{c_k}{\overbrace{\braket{0\ldots00|V|\phi_k}}}\alpha^k \beta^{n-k}}
{\sum_{k=0}^n\underset{d_k}{\underbrace{\braket{0\ldots01|V|\phi_k}}}\alpha^k \beta^{n-k}}\\
&=\frac{\sum_{k=0}^n c_k \alpha^k \beta^{n-k}}{\sum_{k=0}^n d_k \alpha^k \beta^{n-k}}
=\frac{\sum_{k=0}^n c_k z^k}{\sum_{k=0}^n d_k z^k}
\end{split}
\end{equation}

\pagebreak

We present a linear algebraic argument showing that for any rational map of degree $\geq 2$ there is a suitable unitary $V$, thus we need to find a unitary for any coefficients $c_k,d_k$ describing a rational map. 

First let us introduce some vector $\ket{\omega}=(\ket{\underset{n-2 \text{ zeros}}{\underbrace{0\ldots0}}10}-\ket{\underset{n-2 \text{ zeros}}{\underbrace{0\ldots0}}01})/\sqrt{2}$ which is orthogonal to all the $\ket{\phi_k}$ vectors. We set $\tilde{v_0}=\sum_{k=0}^n c_k \frac{\ket{\phi_k}}{\sqrt{n\choose k}} + x\cdot\ket{\omega}$ and $\tilde{v_1}=\sum_{k=0}^n d_k \frac{\ket{\phi_k}}{\sqrt{n\choose k}} + y\cdot\ket{\omega}$. Then by choosing $x,y \in \mathbb{C}$ appropriately we can always satisfy the equalities:

\begin{equation}
\label{eq:ortho}
y^*x=-\sum_{k=0}^n c_k^*d_k
\Leftrightarrow \sum_{k=0}^n c_k^*d_k+y^*x= 0
\Leftrightarrow \braket{\tilde{v_0}|\tilde{v_1}} = 0 
\end{equation}
\begin{equation}
\begin{split}
\label{eq:normed}
&|x|^2-|y|^2=\sum_{k=0}^n (|d_k|^2-|c_k|^2)\\
&\Leftrightarrow \sum_{k=0}^n |c_k|^2+|x|^2=\sum_{k=0}^n |d_k|^2+|y|^2
\Leftrightarrow |\tilde{v_0}|=|\tilde{v_1}| 
\end{split}
\end{equation}

Finally setting $\bra{\underset{n-1 \text{ zeros}}{\underbrace{0\cdots0}}0}V:=\bra{\tilde{v_0}}/\braket{\tilde{v_0}|\tilde{v_0}}$
and $\bra{\underset{n-1 \text{ zeros}}{\underbrace{0\ldots0}}1}V:=\bra{\tilde{v_1}}/\braket{\tilde{v_1}|\tilde{v_1}}$
satisfies \eqref{eq:nmap} and also \eqref{eq:ortho},\eqref{eq:normed} so we can extend $V$ to a full $n$ qubit unitary by defining the remaining $2^n-2$ orthonormal rows arbitrarily.

Note that if $\sum_{k=0}^n c_k z^k$ and $\sum_{k=0}^n d_k z^k$ has no common roots, than the probability that the process succeeds is greater than some $p$ probability regardless the state $z$. The probability cannot be zero since there is no common root, and it follows from compactness that is cannot approach zero for a fixed map. However, depending on the map, this lower bound may be arbitrarily low.

Thus using the above defined $V$ unitary we can implement the rational function $ \frac{\sum_{k=0}^n c_k z^k}{\sum_{k=0}^n d_k z^k} $ where as before this means a transformation 
$$(\alpha\ket{0}+\beta\ket{1})\rightarrow N\left(\sum_{k=0}^n c_k \alpha^k\beta^{n-k}\ket{0}+\sum_{k=0}^n d_k \alpha^k\beta^{n-k}\ket{1}\right)$$
where $z=\alpha/\beta$ and $N$ is a norming factor. 



\mywordcount

\end{document}